\documentclass[fleqn,10pt]{wlscirep}
\usepackage[utf8]{inputenc}
\usepackage[T1]{fontenc}
\usepackage{graphicx}
\usepackage{dcolumn}
\usepackage{bm}
\usepackage{amssymb}
\usepackage{amsmath}
\usepackage{amsfonts}
\usepackage[]{graphicx}
\def\beq{\begin{eqnarray}}
\def\eeq{\end{eqnarray}}

\begin{document}

\title{Comment on ``Energy–speed relationship of quantum
particles challenges Bohmian mechanics''}

\author[1,*]{Aur\'elien Drezet}
\author[2]{Dustin Lazarovici}
\author[2]{Bernard Michael Nabet}

\affil[1]{Institut N\'eel, UPR 2940, CNRS-Universit\'e Joseph Fourier, 25, rue des Martyrs, 38000 Grenoble, France}
\affil[2]{Technion--Israel Institute of Technology, Humanities and Arts Department, Haifa, 3200003, Israel}

\affil[*]{aurelien.drezet@neel.cnrs.fr}



\date{\today}

\begin{abstract}
No, it does not. 
\end{abstract}
\maketitle
\section{Introduction}

\indent In their recent paper published in Nature \cite{Nature}, Sharaglazova et al. report an optical microcavity experiment yielding an ``energy-speed relationship'' for quantum particles in evanescent states, which they infer from the observed population transfer between two coupled waveguides. The authors argue that their findings challenge the validity of Bohmian particle dynamics because, according to the Bohmian guiding equation, the velocities in the classically forbidden region would be zero. In this note, we explain why this claim is false and the experimental findings are in perfect agreement with Bohmian mechanics. In contrast to other recent replies\cite{nikolic2025, wang.etal2025}, our analysis relies solely on the standard Bohmian guidance equation for single particles.

Before turning to more technical points, we highlight several general reasons why the experiment cannot contradict Bohmian mechanics -- unless it also conflicted with standard quantum theory.
First and foremost, there are general results establishing the empirical equivalence of Bohmian mechanics and standard quantum mechanics\cite{durr.etal1992, durr.etal2004}. Although the authors reference recent debates about the scope of these results, these do not appear to be relevant to the present experiment. 

Indeed, and more specifically, the experiment's key findings are based on an operational notion of (non-directional) particle speeds derived from the relative population density of particles in the main and auxiliary waveguides: 
\begin{equation}\label{popratio} \rho_a = \frac{\lvert \psi_a\rvert^2}{\lvert \psi_a\rvert^2 + \lvert \psi_m\rvert^2} \end{equation}
Bohmian mechanics predicts the same wave functions and Born distributions as standard quantum mechanics, hence also the same ``speeds'' in the sense defined by Sharaglazova et al. 

Third, the Bohmian velocity field $v^\psi$, determined by the standard guiding equation, is related to the quantum flux  $j^\psi$ via $ j^\psi = \lvert \psi\rvert^2 v^\psi$, which satisfies the continuity equation $\partial_t \lvert \psi\rvert^2 = - \nabla \cdot j^\psi $ as a direct consequence of the Schr\"odinger equation. If the flux (i.e., probability current) through the classically forbidden region were truly zero, as the authors imply, it would be puzzling -- even from the standpoint of standard quantum mechanics -- how any population transfer between the waveguides could occur. The experiment could even be described entirely in the optical domain, where $j^\psi$ can be identified with the energy flux in the sense of Poynting's theorem. From these considerations alone, it is clear that a vanishing flux would be incompatible with the observation of any signal in the experiment.

We note that there are open foundational questions regarding the treatment of photons in Bohmian mechanics\cite{tumulka2018}, but we do not believe these are essential to the interpretation of the present experiment. De Broglie–Bohm-type trajectories are successfully used in optics\cite{sanz.miret-artes2012} and the experiment implements a paraxial photonic analogue of the Schrödinger equation, which is a well-established method for simulating massive quantum particles\cite{longhi2009}.

\section{The Fallacy in the Argument}\label{sec:fallacy}
The claim that the Bohmian velocities are zero in the classically forbidden region is based on the fact that the stationary scattering states for a step potential can be chosen real-valued in the evanescent region. The corresponding Bohmian velocity field, being proportional to the phase gradient, then vanishes. This is also evident from the equivalent expression
\begin{equation} v^\psi = \frac{\hbar}{m} \mathrm{Im} \frac{\nabla \psi}{\psi}, \end{equation} where $\mathrm{Im}$ denotes the imaginary part.

By design, the experiment probes population distributions in a quasi-stationary regime (the duration of the laser pulse, $26 \mathrm{ns}$, being much longer than the characteristic timescales of the cavity dynamics). A corresponding theoretical treatment in the stationary regime is therefore justified \emph{if} one is only concerned with the observed population densities. But this amounts to considering a quasi-equilibrium situation after Bohmian trajectories have already penetrated the classically forbidden region as a result of earlier transient dynamics. 

If one wants to describe the Bohmian dynamics for particles tunneling through the barrier and moving in the evanescent regime, a stationary treatment is inadequate, and one must consider the time-dependent scattering of the wave packet at the potential step. Taking into account that this wave packet (corresponding to the coherent laser pulse propagating through the main waveguide) has a narrow but non-zero spectral width, the phase gradient does not vanish at the potential barrier. The transmitted component enters the forbidden region with non-zero flux, corresponding to a non-zero Bohmian velocity field.\cite{leavens1990, norsen2013} For a semi-infinite step potential, the trajectories will eventually turn around and leave the classically forbidden region, consistent with the total reflection of the incident waves. 

A detailed discussion of this scattering problem, while instructive, is beyond the scope of this note. A simple calculation, estimating Bohmian velocities in the evanescent region, may proceed as follows. We consider a wave packet of the form

\begin{eqnarray}
\psi(x, t) \propto \int \mathrm{d}E g(E) e^{-iEt} e^{-k_E x},\label{5}
\end{eqnarray} 
where $k_E=\sqrt{2m(V_0-E+m)}\in\mathbb{R}^+$ and we choose
\begin{eqnarray}
g(E) \propto e^{-\frac{(E-E_0)^2}{4\sigma^2}}
\end{eqnarray}  
as a Gaussian with standard deviation $\sqrt{2}\sigma$ around the dominant energy $E_0$. For ease of notation, we work in units with $\hbar =1$ and $c=1$ ($c$ being the speed of light in the resonator medium), and omit normalization factors that cancel out when we consider the Bohmian velocities. 
Expanding $k_E$ to first order around the dominant value  $k_{0}$, i.e.,  $k_E\simeq k_{0}+ \frac{\partial k_E}{\partial E}\bigl\vert_{E_0} (E-E_0)$, we obtain a field of the form
\begin{eqnarray}
\psi(x, t) \propto  e^{\sigma^2(it -\frac{mx}{k_{0}})^2}  e^{-iE_0 t}e^{-k_{0}x}\label{6},
 \end{eqnarray} 
 i.e., a Gaussian envelope modulated by exponential decay.
The corresponding Bohmian velocity field is 
\begin{equation}\label{eq6} v^\psi(x, t) = \frac{1}{m} \mathrm{Im} \frac{\partial_x \psi(x, t)}{\psi(x,t)} = -\frac{2 \sigma^2}{k_0}t.\end{equation}
The estimated travel distance over the pulse duration $\tau \sim \frac{1}{\sigma}$ is comparable to the characteristic decay length $L = \frac{1}{k_0}$ of the evanescent wave. 

Note that the velocity changes sign (in our approximation, always at $t=0$, since we considered only evanescent modes in \eqref{5}, ignoring penetrating tails of the incident wave packet and the phase shift at the barrier). This corresponds to essentially parabolic trajectories, which enter the classically forbidden region before turning around and leaving it. To efficiently image photons associated with a population in the evanescent zone, as in the experiment, another effect is relevant, which we discuss next. 

\section{Radiative Leakage}\label{sec:leakage}
Another factor ignored in the analysis of Sharaglazova et al. is radiative leakage, which should be taken into account even when working in the stationary regime. According to the experimentalists, the mode confined in the microcavity has a typical lifetime of $\Gamma^{-1}=270 \, \mathrm{ps}$ due to losses by the mirrors along the optical axis. Notably, this radiative loss is what enables the imaging of the intracavity population in the first place.\cite{drezet2013imaging}

A standard treatment (in terms of steady states) starts by considering the eigenmodes of the planar cavity in the orthogonal $z$-direction and adding a negative imaginary part to the wave vector $k_z\rightarrow k_z -i\frac{\Gamma}{2}$. This corresponds to a growing mode along the $z$ axis in accordance with the theory of radiative leakage.\cite{drezet2008leakage} In the  $x-y$ plane, this translates into an effective dissipative potential $V(x,y)\rightarrow V(x,y)-i\frac{\Gamma}{2}$, which introduces an additional complex phase factor also in the evanescent regime. Even the corresponding stationary states then have a non-zero Bohmian velocity proportional to $\Gamma$. 

Since, in the present experiment, the coherence time of $\Gamma^{-1} = 270\, \mathrm{ps}$ is much shorter than the temporal width of the pulse ($\sigma^{-1} = 26\,\mathrm{ns}$),  the contribution to the Bohmian velocity from radiative leakage is, in fact, about a hundred times larger than the correction obtained above from the spectral width of the pulse. A more detailed model with quantitative estimates is provided in the Appendix. Figure \ref{fig:1} shows numerical simulations of the resulting Bohmian trajectories, which are fully consistent with the observation of particles in both waveguides. 

\begin{figure}[htbp]
    \centering

        \centering
        \includegraphics[width=0.75\linewidth]{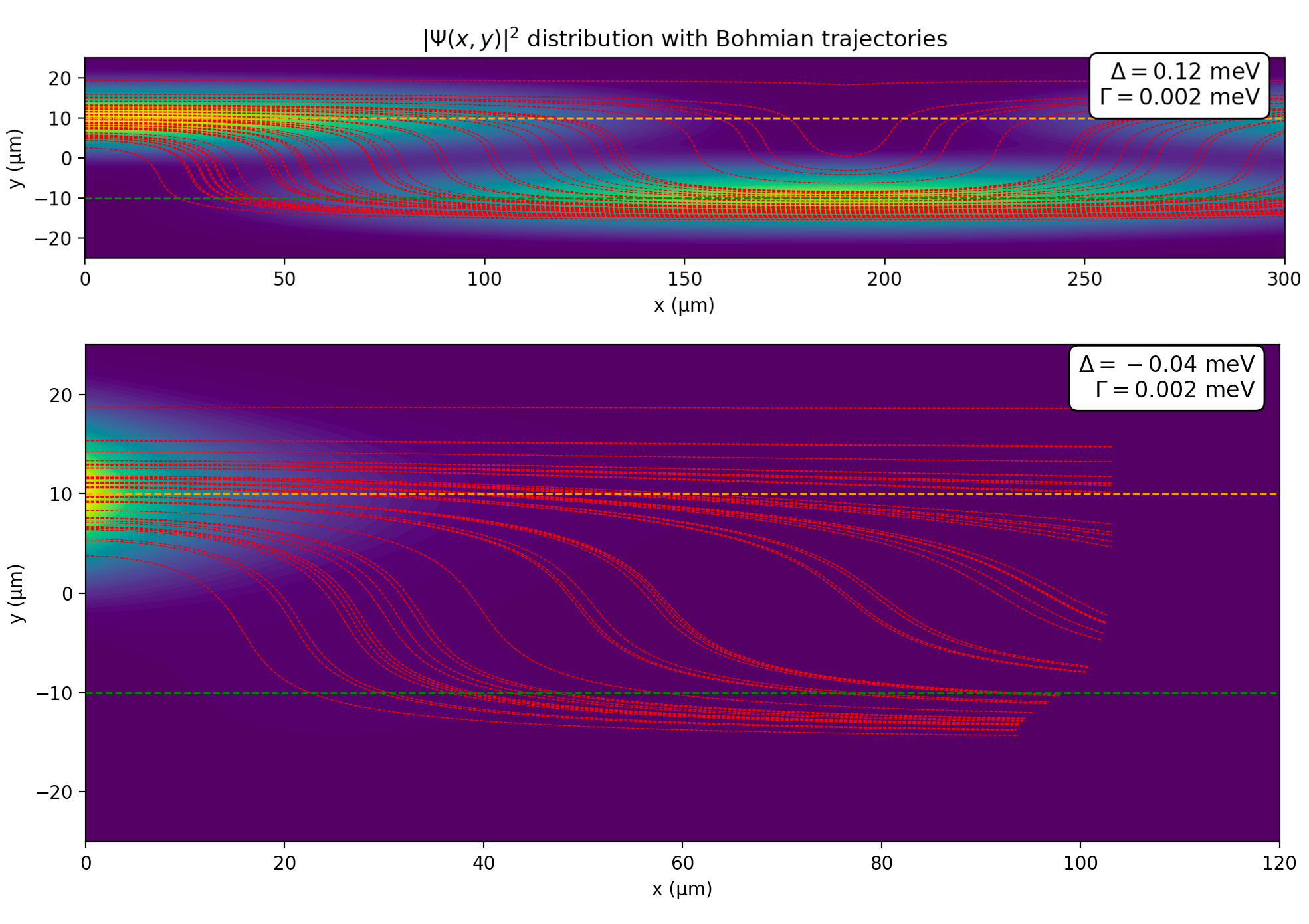}
    
    \caption{Plots of Bohmian trajectories (dotted red lines) in the propagative (top) and evanescent (bottom) regime, extracted from the model with leakage described in the Appendix. $\Delta$ is the kinetic energy offset. The two waveguides are centered at $\pm 10\, \mathrm{\mu m }$. Trajectories leaking out in $z$-direction are not shown, so the density of plotted trajectories does not reflect the $|\Psi|^2$-distribution, which is indicated by the density plot in the background.} 
    \label{fig:1}
\end{figure}

\section{What ``Speeds'' does the Experiment actually Measure?}

While the claim that Bohmian mechanics predicts zero velocity -- and hence infinite dwell time -- in the classically forbidden region is clearly false, one might still wonder about the quantitative discrepancy between the relatively small Bohmian velocities and the speeds on the order of $10^3~\mathrm{km\,s^{-1}}$ reported by Sharaglazova et al. 

The short answer is that the latter is a different quantity, obtained by fitting the observed relative population density \eqref{popratio} to the expression
\begin{equation}\label{vdef}
    \rho_a(x) =: \sin^2(J_0x/v) \approx (J_0 x/v)^2,
\end{equation}
where $J_0$ is the coupling constant between the two waveguides. In the propagative regime, the result corresponds to the velocity of the wave packet along the guides (and hence also to typical Bohmian velocities), but this relationship does not extend to the evanescent regime. We leave it to the reader to judge whether the proposed notion of ``speed'' is still well-motivated for the evanescent regime, based on the discussions in the pertinent publications \cite{Nature, klaers.etal2023}. Here, we only point out that if the coupling between the parallel waveguides is modeled by a double-well potential along the $y$-axis (as suggested, though not explicitly implemented, in the paper), the marginal distributions $\lvert \psi_a(x)\rvert^2$ and  $\lvert \psi_m(x)\rvert^2$ in the stationary evanescent regime are found to be governed not by a sustained lateral flux, but by the differential spatial decay of a symmetric and anti-symmetric mode -- corresponding to the ground and first excited states of the double-well potential with energy levels $E_{\pm}= E_0\pm J_0$. (See Appendix below.) 
This differential attenuation -- the symmetric ground state decaying more steeply in $x$-direction than the excited anti-symmetric state -- is essentially what $v$ quantifies in the evanescent case and why it increases as the kinetic energy offset $\Delta:= E - m - V_0 + \hbar J_0$ becomes more negative. 

Thus, the suggested connection to tunneling times is far from evident, and any relation to actual particle velocities -- as described by Bohmian mechanics -- appears tenuous.  

\section*{Acknowledgements}
This work was supported in part by the ISRAEL SCIENCE FOUNDATION (grant No. 1597/23 to D.L.). We thank Jean Bricmont, Sheldon Goldstein, Nino Zanghì, and Roderich Tumulka for helpful discussions.

\appendix
\section*{Appendix: Waveguide Model with Radiative Leakage}\label{appendix}
To further substantiate the points made in our discussion, we expand the theoretical treatment offered by Sharaglazova et al. to include a) the shape of the wave function in the $x-y$ plane and b) the effects of radiative leakage. 
A simple approach to a) is based on the hybridization method commonly used in quantum chemistry and waveguide theory. 
\subsection*{Waveguide model and mode hybridization}

We assume a double well potential $V(y)$ with two energy levels $E_{\pm}=E_0\pm J_0$, more specifically, $E_+=V_0$ and $E_-=V_0-2J_0$. (We work in units where $\hbar = c = 1$, and ignore the refractive index of the resonator medium, which we estimate to be about $n \approx 1.4$ in the reported experiments.) The corresponding eigenstates satisfy 
\begin{equation}
E_\pm \, \Phi_\pm(y) = \left[ -\frac{1}{2m} \frac{\partial^2}{\partial y^2} + V(y) \right] \Phi_\pm(y)
\end{equation}
 and describe antisymmetric and symmetric modes  $\Phi_{\pm}(y)=\frac{\Phi_m(y)\pm\Phi_a(y)} {\sqrt{2}}$, where $\Phi_m$ and $\Phi_a$ have opposite signs and are localized in the main and auxiliary waveguides, respectively. Note that, with our conventions, $\Phi_-$ denotes the symmetric ground state with energy $E_-$.

Since at $x=0$ the wave function should be concentrated in the main guide only, we can make the following ansatz for the wave function in the $x-y$ plane:
\begin{eqnarray}\label{2dfield1}
    \Psi(x,y)=Ae^{ik_2x}[\cos{(k_1x)}\Phi_m(y)-i\sin{(k_1x)}\Phi_a(y)],
\end{eqnarray}
with $k_1=\frac{k_--k_+}{2}$, $k_2=\frac{k_-+k_+}{2}$, $k_\pm=\sqrt{2m(E-E_\pm-m)}$, implying $k_1k_2=mJ_0$ and $E=\frac{k_1^2+k_2^2}{2m}+m +V_0-J_0.$

The relevant regime corresponds to $\Delta = E - m - V_0 + J_0 
<-J_0$, for which the wave vectors are purely imaginary, $k_1=i\kappa_1$, $k_2=i\kappa_2$, and \eqref{2dfield1} becomes
\begin{equation}\label{2dfield}
    \Psi(x,y)=Ae^{-\kappa_2 x}[\cosh{(\kappa_1x)}\Phi_m(y)+\sinh{(\kappa_1x)}\Phi_a(y)].
\end{equation} 
\subsection*{Interpretation of operational speed}
\noindent In this stationary approximation, we have 
\begin{equation} \rho_a(x) := \frac{\rvert \sinh(\kappa_1 x)\lvert^2}{\rvert \sinh(\kappa_1x) \rvert^2 + \lvert \cosh(\kappa_1x) \rvert^2 } =  \frac{\rvert \sinh(\kappa_1 x)\lvert^2}{1 + 2 \lvert \sinh(\kappa_1x) \rvert^2 }, 
\end{equation}
and thus $\rho_a(x) \approx (\kappa_1x)^2$ for small $x$. Comparing with \eqref{vdef}, we obtain
\begin{equation} 
v = \frac{J_0}{\kappa_1} = \frac{2J_0}{\kappa_- - \kappa_+} = \frac{2J_0}{\sqrt{2m(- \Delta + J_0)}- \sqrt{2m(- \Delta - J_0)}}.
\end{equation}
Hence, if $v$ is fit according to the relationship used by Sharaglazova et al., it effectively quantifies the differential attenuation of the symmetric and antisymmetric modes given by the decay constants $\kappa_+$ and $\kappa_-$, respectively. 

\subsection*{Corrections from radiative leakage}
Returning to \eqref{2dfield}, we see that if $\Phi_{m/a}$ are real-valued, the wave function $\Psi(x,y)$ entails a vanishing Bohmian velocity in the $x-y$ plane, as claimed by Sharaglazova et al. However, we already explained why this steady-state approximation cannot be taken at face value. To see Bohmian motion, we could take the spectral width of the pulse into account, as discussed in Section \ref{sec:fallacy}. But here, we want to focus on the (dominant) effect of radiative leakage, as mentioned in Section \ref{sec:leakage}.  

The usual treatment method consists of first taking the eigenmodes of the cavity in the vertical direction $z$, which amounts to adding a negative imaginary part to the wave vector $k_z\rightarrow \frac{q\pi}{D_0}-i\frac{\Gamma}{2}$, where $D_0$ is the typical height of the cavity (here, $D_0 \approx 15 \, \mathrm{\mu m}$) and $q$ an integer. In the  $x-y$ plane, this translates into a dissipative potential $V(x,y)\rightarrow V(x,y)-i\frac{\Gamma}{2}$ which modifies the wave equation for $\Psi(x,y)$:
\begin{eqnarray}
    E\Psi(x,y)=(m+V(x,y)-i\frac{\Gamma}{2})\Psi(x,y)-\frac{\partial_x^2}{2m}\Psi(x,y)-\frac{\partial_y^2}{2m}\Psi(x,y)
\end{eqnarray}
Assuming the same profile  $V(y)$ for the  confining potential and keeping the same transverse eigenmodes   $\Phi_{\pm}(y)$ the full wave function takes the same form as before,
\begin{eqnarray}
    \Psi(x,y)=Ae^{ik_2x}[\cos{(k_1x)}\Phi_m(y)-i\sin{(k_1x)}\Phi_a(y)],
\end{eqnarray} with  $k_1=\frac{k_--k_+}{2}$, $k_2=\frac{k_-+k_+}{2}$, but now $k_\pm=\sqrt{2m(E-E_\pm+i\frac{\Gamma}{2}-m)}$ leading to generally complex wave vectors. 
The corresponding Bohmian velocity field reads:
\begin{equation}\begin{split}
  &v^\psi_{x}(x,y)=\frac{1}{m}\mathrm{Im}\left[\frac{\partial_x\Psi(x,y)}{\Psi(x,y)}\right]=\mathrm{Re}\left[\frac{k_2}{m}\right]+\mathrm{Im}\left[\frac{k_1(-\sin{(k_1x)}\Phi_m(y)-i\cos{(k_1x)}\Phi_a(y))}{
    m\left(\cos{(k_1x)}\Phi_m(y)-i\sin{(k_1x)}\Phi_a(y)\right)}\right]\\
 &v^\psi_{y}(x,y)=\frac{1}{m}\mathrm{Im}\left[\frac{\partial_y\Psi(x,y)}{\Psi(x,y)}\right]=\mathrm{Im}\left[\frac{\cos{(k_1x)}\partial_y\Phi_m(y)-i\sin{(k_1x)}\partial_y\Phi_a(y)}{
    m\left(\cos{(k_1x)}\Phi_m(y)-i\sin{(k_1x)}\Phi_a(y)\right)}\right]
\end{split}\end{equation}
\noindent In particular, we have at $x=0$ (the start of the step potential): $v^\psi_{x}(0,y)=\mathrm{Re}\left[\frac{k_2}{m}\right]-\mathrm{Re}\left[\frac{k_1}{m}\frac{\Phi_a(y)}{
    \Phi_m(y)}\right]$ and  $v^\psi_{y}(0,y)=0$. 
    
Importantly, for $\Gamma/m\ll1$, we now have 
\begin{eqnarray}
    k_{\pm}\simeq k_{\pm}^{(0)}+i\frac{m\Gamma}{2 k_{\pm}^{(0)}}
\end{eqnarray}
where $ k_{\pm}^{(0)}$ are the previous values of the wavevectors without damping $\Gamma$. Thus, the Bohmian velocity is non-vanishing even in the evanescent regime, where $ k_{\pm}^{(0)}$ are purely imaginary numbers. 

For $k_2 \gg k_1$, we can further approximate 
\begin{eqnarray}
     v^\psi_{x}\approx \mathrm{Re}\left[\frac{k_2}{m}\right]\approx \frac{\Gamma}{2m}\frac{1}{v_\Delta}
\end{eqnarray}
where $v_\Delta=\sqrt{-2\Delta / m}$ interpolates the speed $v$ measured by Sharaglazova et al. With $v_\Delta \approx 2000\, \mathrm{km\,s^{-1}}$ i.e., in our  units $v_\Delta \approx 10^{-2}$, and $\frac{\Gamma}{2m} \approx 10^{-6} $, we find  $v^\psi_{x}\approx 
10^{-4} \approx  30\,\mathrm{km\,s^{-1}}$, in good agreement with the value measured by the authors using an interferometric device.  
We emphasize that this is only a rough estimate, although the corresponding experimental measurement is also subject to significant uncertainty.

\bibliography{references}
\end{document}